\documentclass[reprint, superscriptaddress, amsmath, amssymb, aps, floatfix]{revtex4-2}

\usepackage[columnwise,running]{lineno}
\usepackage{orcidlink}
\usepackage{hyperref}
\usepackage{physics}
\usepackage{ragged2e}

\usepackage{amsmath}
\usepackage{amssymb}
\usepackage{amsfonts}
\usepackage{amstext}
\usepackage{amsthm}
\usepackage{bbm}

\usepackage{graphicx}
\usepackage{dcolumn}
\usepackage{bm}

\usepackage[utf8]{inputenc}
\usepackage[T1]{fontenc}

\usepackage{braket}
\usepackage{siunitx}
\usepackage{xcolor}

\usepackage{booktabs}
\usepackage{multirow}

\usepackage[normalem]{ulem}

\def\ddt{\dd t}

\begin{document}

\title{Deterministic quantum master equation for non-Markovian signal processing}
\author{Guilherme De Sousa\ \orcidlink{0000-0002-8529-5439}}
\email{guilherme2.desousa@gmail.com}

\author{Diogo O. Soares-Pinto\ \orcidlink{0000-0002-4293-6144}}
\email{dosp@ifsc.usp.br}

\affiliation{Instituto de Física de São Carlos$,$ Universidade de São Paulo$,$ IFSC-USP, 13560-970 São Carlos$,$ SP$,$ Brasil}
\date{\today}

\begin{abstract}
    In this work, we derive a deterministic master equation to model a general, possibly non-Markovian, feedback.
    The master equation describes a system with a general evolution and measurement operation, with feedback being applied in terms of signal processing.
    The feedback signal has an arbitrary structure with dimensionality that indicates the degree of non-Markovianity of the information processing.
    We present examples to illustrate how such a master equation can be used to model systems with memory feedback and non-trivial frequency dependence.
\end{abstract}

\maketitle


\section{Introduction}\label{sec:introduction}

Feedback control is ubiquitous in the development of next-generation quantum technologies.
{In quantum optomechanics~\cite{Hopkins2003,Guo2019,Manikandan2023} and trapped-ion platforms~\cite{Bushev2006}, feedback protocols have been employed to cool mechanical and motional degrees of freedom.
Feedback-based control can also play an important role in quantum thermodynamics~\cite{Cao2009}, where measurements and adaptive protocols can enhance work extraction and engine performance~\cite{Manikandan2022,Liu2025,Brandner2015,Potts2019}.
More recently, non-Markovian feedback schemes have attracted growing interest in quantum information processing and error correction~\cite{Puviani2025}.}

In the past decades, various models to study quantum feedback systems were developed, from single and repeated measurements~\cite{Sagawa2008,Funo2013,Lewalle2020} to diffusive and continuous measurement models~\cite{Yada2022,AnnbyAndersson2022,Tilloy2024,Rosal2025}.
Some approaches rely on studying the stochastic trajectories and need a statistical treatment to calculate ensemble quantities~\cite{Jacobs2006,Horowitz2012,Manikandan2022}.
Conversely, deterministic equations can be derived by considering the evolution of quantum states consistent with the ensemble of trajectories~\cite{Diosi2014,Diosi2023,Layton2024,Tilloy2024,Rosal2025,DeSousa2025_2}.
A recent formulation by Rosal \emph{et al}~\cite{Rosal2025}, generalizes most of these approaches by constructing an ensemble average, deterministic master equation that can recover results from discrete or continuous measurements.
This deterministic master equation is suitable for Markovian signal processing and is directly translated to well-known master equation formulations.

While stochastic trajectory approaches can incorporate memory effects, a general deterministic description remains lacking.
This limits analytical understanding and efficient modeling of feedback protocols with structured temporal correlations.
Here, we provide a deterministic formulation that captures arbitrary non-Markovian signal processing through a finite-dimensional Markovian embedding.
Throughout, we characterize non-Markovianity by a general {signal that evolves with a dependency on past values}
\begin{equation}
    \label{eq:general_non_markovian_signal}
    s_{n+1} = g_{n+1}(x_{n+1},s_n,s_{n-1},\dotsc,s_{n-T}).
\end{equation}
{The non-Markovian signal $s_n$ is then used to apply feedback into the quantum system, e.g., by changing the Hamiltonian using $H=H(s_n)$.
The evolution is induced by the function $g_n$ representing the controller dynamics, and $x_{n+1}$ denotes the measurement outcome resulting from the quantum measurement step during the feedback loop.}
{This notion of non-Markovianity should not be confused with quantum non-Markovianity of the induced dynamical map in the sense of BLP or RHP \cite{Breuer2009,Rivas2010,Breuer2016}.}

The main result of this paper is to show that the evolution of the feedback-resolved state with general non-Markovian feedback follows
\begin{equation}
    \label{eq:deterministic_ME}
    \varrho_{n+1}(\vec{y}) = \sum_{x',\vec{y}'} \delta_{\vec{y},\vec{f}_{n+1}(x',\vec{y}')} \mathcal{M}_{x'}(\vec{y}') \varrho_n(\vec{y}'){.}
\end{equation}
{Here, $\varrho({\vec y})$ denotes the feedback-resolved joint state that encodes information about the quantum state and the probability distribution of the signal vector $\vec{y}$.
The signal $\vec{y}_n$ is a Markovian-embedded version of the signal $s_n$, and evolves under the dynamics generated by the function $\vec{f}_n$; the values $x$ represent measurement outcomes.
Finally, the operator $\mathcal{M}_x({\vec y})$ encodes the feedback-dependent quantum evolution acting on the feedback-resolved density matrix.}


\begin{figure*}
    \centering
    \includegraphics[width=0.8\linewidth]{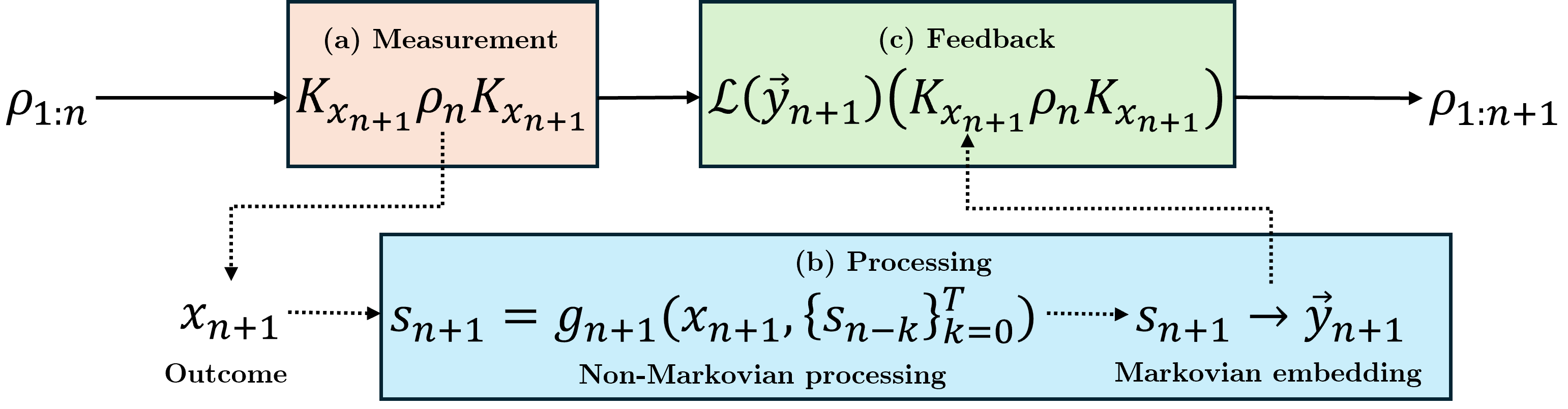}
    \caption{Steps for evolving the conditional quantum state from one observation at $t_n$ to the next observation at $t_{n+1}$. The sequence involves (a) measuring using the Kraus operators, (b) collecting and processing the information using non-Markovian feedback rule, and (c) applying feedback based on the Markovian embedded high-dimensional signal.}
    \label{fig:feedback-steps}
\end{figure*}

In this formulation, the signal is characterized by a high-dimensional vector $\vec{y}$, whose components encode information about the instantaneous signal $s_n$ and transformations of past values $s_{1},\dotsc,s_{n-T}$.
The dimensionality of $\vec{y}$ indicates how many additional signals are necessary to resolve the non-Markovianity.
The function $\vec{f}_n$ is a vector function that evolves the high-dimensional signal $\vec{y}_n$ to the next step, and it is related to the non-Markovian {controller dynamics} in Eq.~\eqref{eq:general_non_markovian_signal}.
The increase of the signal's dimension can be mapped into the general form of hybrid master equations, which can be used to model general filtering operations~\cite{Tilloy2024,DeSousa2025_2}.

This paper is structured as follows: in Sec.~\ref{sec:derivation} we outline the derivation of the main result in Eq.~\eqref{eq:deterministic_ME}. Section~\ref{sec:non-markovian} shows the connection between the non-Markovian modeling of the signal processing and the deterministic master equation. We finish with discussion and conclusion in Sec.~\ref{sec:outlook}

\section{Outline of the derivation}\label{sec:derivation}
Let's define our system and the steps necessary to apply feedback: we consider a quantum system with discrete or continuous state space, with the quantum state described by a density matrix $\rho_n \equiv \rho(t_n)$, with $t_n$ corresponding to the observation times when measurement and feedback take place.
At each observation step, we (a) measure, (b) process the information, and (c) change the evolution based on the processed signal, thus applying feedback --- see Fig.~\ref{fig:feedback-steps}.

The measurement is modeled using a set $\{ K_x \}$ of Kraus operators that satisfy $\sum_x K_x^\dagger K_x = \mathbbm{1}$, with $x$ being the measurement outcome index (continuous or discrete), and the probability associate with the measurement is given by Born rule $P(x_{n+1}) = \tr{K_{x_{n+1}} \rho_n K_{x_{n+1}}^\dagger}$.
The post measurement state is given by{~\cite{HPBreuer2007}:}
\begin{equation}
    \rho_{n+1,{\rm post~measurement}} = \frac{K_{x_{n+1}} \rho_n K_{x_{n+1}}^\dagger}{P(x_{n+1})}.
\end{equation}
It is worthwhile to note that this evolution is stochastic because the measurement outcome $x_{n+1}$ is a stochastic process.
Thus, we will refer to the evolution of the density matrix $\rho_n$ to be \textit{conditioned} on the sequence $x_{1:n} \equiv \{ x_1,\cdots, x_n \}$.

Based on the measurement outcome $x_{n+1}$, the experimentalists implement a signal processing step, consisting of a signal $s_n \equiv s(t_n)$ from which feedback is applied.
The evolution of the signal $s_n$ is deterministic and is given by the update rule in Eq.~\eqref{eq:general_non_markovian_signal}.
If the update function only depends on $g_n(x_{n+1},s_n)$ the processing is said to be Markovian, and the results follow Ref.~\cite{Rosal2025}.
When the update function depends on past values $\{ s_{n-1},\dotsc, s_{n-T} \}$, the signal processing is non-Markovian, thus a \textit{Markovian embedding} is necessary.

The Markovian embedding in this context consists of promoting the scalar signal to a high-dimensional vector $s_n \rightarrow \vec{y}_n$ with appropriate dimension to resolve the non-Markovianity of the {controller} function $g_n$~{\cite{Kailath1980,Chen2013}}.
{This procedure is equivalent to transforming a $n-$order differential equation into a set of $n$ first-order differential equations.}
This process is model-dependent and is illustrated in Sec.~\ref{sec:non-markovian}.
Using the signal function $s_n$ (or its embedded version $\vec{y}_n$), we can apply feedback.
In the embedding space, the update function is given by
\begin{equation}
    \vec{y}_{n+1} = f_{n+1}(x_{n+1},\vec{y}_n).
\end{equation}

Feedback is modeled as a generic parameter from which the time evolution can be changed.
The time evolution is given by a general CPTP channel
\begin{equation}
    \rho_{n+1} = \mathcal{L}_{n+1}(\vec{y}_{n+1}) \rho_{n+1,{\rm post~measurement}}.
\end{equation}
Combining with the measurement step, we can write the evolution of our conditioned density matrix at the next observation time $t_{n+1}$ as
\begin{equation}
    \label{eq:rho_conditioned_evolution}
    \rho_{n+1} = \mathcal{L}_{n+1}(\vec{y}_{n+1})\left[ \frac{K_{x_{n+1}} \rho_n K_{x_{n+1}}^\dagger}{P(x_{n+1})} \right] \equiv \frac{M_{x_{n+1}}(\vec{y}_n) \rho_n}{P(x_{n+1})},
\end{equation}
where $M_{x_{n+1}}(\vec{y}_n)$ defines a quantum instrument~\cite{Manzano2020,Milz2021}.

The final step to derive Eq.~\eqref{eq:deterministic_ME} is to define an ensemble-resolved density matrix $\varrho_n(\vec{y})$ that averages all possible conditional quantum states $\rho_n$ that are consistent with the signal $\vec{y}_n$ at time $t_n$,
\begin{equation}
    \varrho_n(\vec{y}) = {\mathrm E}\left[ \rho_n \delta_{\vec{y},\vec{y}_n} \right].
\end{equation}
Here, ${\mathrm E}[\cdot]$ denotes the average over the stochastic measurement outcomes $x_{1:n}$. This operator is a joint-distribution between quantum degrees of freedom and measurement-outcome (signal-processed) degrees of freedom.
In Appendix~\ref{sec:app-derivation}, we present a detailed derivation of Eq.~\eqref{eq:deterministic_ME}.
{Note that all steps in the derivation of Eq.~\eqref{eq:deterministic_ME}, satisfy the normalization property of $\sum_{\vec y}\tr{\varrho_n({\vec y})} = 1$.
Meaning that our evolution map is trace preserving, as shown in Appendix~\ref{sec:app-trace}}.



\section{Application to non-Markovian feedback}\label{sec:non-markovian}
Here, we show how one can design a controller function $\vec{f}_n$ to have access to $s_{n-1,\dotsc,n-T}$ in the embedded space $\vec{y}_n$, thus applying a non-Markovian signal processing.
The particular structure of the signal processing $\vec{f}_n$ is model-dependent and can change according to the internal structure of the high-dimensional vector $\vec{y}$.
As a first example, let's consider how one can incorporate a single step from the past into the feedback rule.

\subsection{Feedback rule with momentum}
Let's consider a general {controller dynamics} that only depends on the instantaneous value of the signal $s_n$:
\begin{equation}
    s_{n+1} = s_n + {\delta t v}_{n+1}(x_{n+1}),
\end{equation}
{with $\delta t \equiv t_{n+1}-t_n$.}
A modification of the feedback to include a momentum-like term (which adds memory effects) can be done in the following way:
\begin{equation}
    \label{eq:update_momentum}
    s_{n+1} = s_n + {\delta t}(1-\beta){v}_{n+1}(x_{n+1}) + \beta (s_n - s_{n-1}).
\end{equation}
Here, $\beta$ controls the relative contribution of the momentum (or inertia) of the signal $s_n$ compared to the feedback rule ${v}_n$.

To describe this system in a Markovian way, let's define an auxiliary signal $m_n$, such that
\begin{equation}
\begin{aligned}
    \label{eq:markovian_momentum}
    s_{n+1} &= s_n + {\delta t} m_{n+1}{,}\\
    m_{n+1} &= \beta m_n + (1-\beta){v}_{n+1}(x_{n+1}){.}
\end{aligned}
\end{equation}
The composite signal $\vec{y}_n = (s_n,m_n)^T$ evolves under a Markovian update rule $\vec{y}_{n+1} = \vec{f}_{n+1}(x_{n+1},\vec{y}_n)$.
This particular model of adding momentum into the evolution of the controller signal $s_n$ is closely related to the Nesterov accelerated gradient algorithm used in optimization problems~\cite{Nesterov1983}.

For the case of continuum feedback $\delta t \rightarrow 0$, if we define $\beta {\equiv} 1 - \gamma\delta t$, we can write Eq.~\eqref{eq:markovian_momentum} as:
\begin{equation}
\begin{aligned}
    &{\dd m \equiv m_{n+1}-m_n = \gamma \left( v_{n+1}(x_{n+1}) - m \right)\dd t}, \\
    &{\dd s \equiv s_{n+1}-s_n = m\ddt}.
\end{aligned}
\end{equation}
{Whose solution can be found analytically~\cite{Hairer1993,Butcher2016}:}
\begin{equation}
\label{eq:momentum-feedback-filtered-signals}
    \begin{aligned}
        m(t) &= \gamma \int_{-\infty}^t~{\dd z}~e^{-\gamma(t-{z})} {v}(x_{{z}}), \\
        s(t) &= \int_{-\infty}^t~{\dd z}~m({z}).
    \end{aligned}
\end{equation}
Or equivalently,
\begin{equation}
    s(t) = \gamma \int_{-\infty}^t~{\dd z}~\left(1-e^{-\gamma(t-{z})}\right) {v}(x_{{z}}),
\end{equation}
showing that the inclusion of momentum is equivalent to a kernel with memory (past values have substantial weight).
{Note that we extended the inferior integration limit to $-\infty$ in the definition of the filtered signals in Eq.~\eqref{eq:momentum-feedback-filtered-signals}.
This can be done for two main reasons: a) for a sufficiently well-behaved function $v(x)$, the exponential kernel in $m(t)$ will regularize the past infinity dependence of the signal; b) instead of adding a new parameter $t_0$ to the definition of the filtered signals, without loss of generality, the feedback function can be thought to have a starting point of the form $v(t)\sim \Theta(t-t_0)$, that automatically accounts for a finite past time dependence.}

\subsection{Feedback rule with $T_{{s}}$ steps in the past}
Now, we turn to the problem of modeling a {controller dynamics} that can depend on $T_{{s}}$ steps in the past, as represented by Eq.~\eqref{eq:general_non_markovian_signal}.
The exact construction from the non-Markovian model to a Markovian embedding using the vectors $\vec{y}_n$ and $\vec{f}_n$ is model-dependent and can vary for different signal processing implementations.
Here, we show one possible implementation that works for a general non-Markovian {signal processing} described by Eq.~\eqref{eq:general_non_markovian_signal}.

Define $m_n^{(k)} = s_{n+1-k} - s_{n-k}$ as a set of auxiliary momentum signals, from which one can write $s_{n-k} = s_n - \sum_j^{k} m_n^{(j)}$.
The evolution of signals can be written as:
\begin{equation}
\label{eq:non-markovian-system}
    \begin{aligned}
        s_{n+1} &= g_{n+1}\left(x_{n+1},s_n,s_n-m_n^{(1)},\dotsc,s_n-\sum_j^{T_s}m_n^{(j)}\right){,}\\
        m_{n+1}^{(1)} &= s_{n+1} - s_n{,} \\
        m_{n+1}^{(2)} &= m^{(1)}_n{,} \\
        &~~\vdots\\
        m^{(T_s)}_{n+1} &= m_{n}^{(T_s-1)}{.} \\
    \end{aligned}
\end{equation}
This system of equations is Markovian because the future values at ${t_{n+1}}$ only depend on quantities evaluated at time ${t_n}$.
The construction is not unique, but it is general, and shows how one can apply non-Markovian feedback using a Markovian embedding into a high-dimensional signal.
Note that the number of equations matches on how many steps in the past that are taken into account.
From the construction in Eq.~\eqref{eq:non-markovian-system}, we can define the vector $\vec{y}_n = (s_n,m_n^{(1)},\dotsc,m_n^{(T_{{s}})})^T$ and the appropriate high-dimensional {controller dynamics} $\vec{f}_n$ to be used in Eq.~\eqref{eq:deterministic_ME}.

This construction also works for continuous measurement.
In the limit $t_{n+1}-t_n \equiv \delta t \rightarrow 0$, one can use Gaussian Kraus operators in Eq.~\eqref{eq:rho_conditioned_evolution} and linear transformation for $g_n$ to recover the results in Refs.~\cite{Tilloy2024,DeSousa2025_2}.

\section{Discussion and outlook}\label{sec:outlook}
\textit{Discussion}---The general deterministic equation for feedback control in Eq.~\eqref{eq:deterministic_ME} can reproduce all previous results in the field for appropriate choices of {controller dynamics} and quantum instruments.
This framework is particularly relevant for quantum control protocols with finite-bandwidth electronics, delayed feedback, or digital filtering, where memory effects {naturally arise.}

{In realistic experiments, finite controller response times and signal-processing introduce temporal correlations in the feedback signal, as routinely observed in superconducting-circuit and optomechanical platforms employing real-time feedback~\cite{Minev2019,Guo2019}.
Delayed coherent feedback generated by optical or microwave propagation lines also provides a natural realization of non-Markovian dynamics in photonic circuits and waveguide QED~\cite{Pichler2016,Regidor2021}.
Within our formalism, these effects are encoded in the dimensionality of the auxiliary signal vector ($\vec{y}_n$), which can be interpreted as the effective memory depth of the controller.}

If the signal $\vec{y}_n \rightarrow s_n$ is a scalar and consequently, the {controller} function is Markovian $g_n \equiv g_n(x_{n+1},s_n)$, we recover the master equation derived in Ref.~\cite{Rosal2025}.
In their previous work, the authors showed how the Markovian feedback version of Eq.~\eqref{eq:deterministic_ME} can recover quantum jump dynamics, diffusion feedback, and the Quantum Fokker-Planck Master Equation~\cite{AnnbyAndersson2022}.
By increasing the dimensionality of the signal processing, our general master equation can also recover the quantum Fokker-Planck master equation with general filtering~\cite{Tilloy2024,DeSousa2025_2}.
This happens for the case where the instruments are described by Gaussian Kraus operators, and the {update} rule is a general linear transformation $\vec{f}_{n+1} = {\rm  M}\vec{y}_n + {\vec b}x_{n+1}$.

Equation~\eqref{eq:non-markovian-system} shows how a general a signal processing with memory can be modeled.
This construction is model-dependent because it is affected by the particular signal processing happening in the experiment.
For the case of the quantum Fokker-Planck master equation with general filtering~\cite{DeSousa2025_2}, the particular choices of the momentum signals defines the particular filtering kernel used in the experiment.
Here, we also showed the trade-off between increasing the memory of the signal processing (by accessing more steps in the past) and the dimensionality of the vector signal $\vec{y}_n$, which is a direct consequence of the Markovian embedding that allows a deterministic equation to be derived.

{It is also important to note that several non-Markovian master-equation approaches have been developed for quantum optics and open quantum systems.
These formulations capture effects such as strong system-reservoir correlations~\cite{Shen2018_2,Shen2020}, dynamics beyond the rotating-wave approximation~\cite{Chang2010,Shen2018}, and delayed coherent feedback~\cite{Li2024}.}
{The present framework is formulated at the level of generalized feedback signal processing and is therefore largely independent of the microscopic origin of the open-system dynamics.
Incorporating ultrastrong-coupling or non-rotating-wave effects may require modified quantum instruments or enlarged signal spaces to account for additional reservoir correlations and coherent backaction.
Exploring these extensions constitutes an interesting direction for future work.}

\textit{Conclusion}---In this work, we extended the results of Ref.~\cite{Rosal2025} to the case of non-Markovian signal processing.
We illustrated how one can incorporate memory effects in signal processing by explicitly constructing a momentum-dependent update rule that has applications in optimization problems.
Then, we show a particular signal processing structure that can implement general non-Markovian processing depending on $T_{{s}}$ steps in the past.
This construction shows that, to have access to $T_{{s}}$ past values of the signal, one needs to increase the dimensionality of the system to $(T_{{s}}+1)$-dimensions.
Similar dimensionality effects were also found in Refs.~\cite{Tilloy2024,DeSousa2025_2}.

We hope our results will add to the far-reaching literature of quantum feedback and to the recent developments of quantum technologies.
The ability to describe quantum feedback in a deterministic way can change how people design and optimize experiments, interact with quantum systems, and even develop new applications to real-world problems.

\textit{Acknowledgment}---The authors thank     Alberto~J.~B.~Rosal for fruitful discussions.
This work was funded by the São Paulo Research Foundation (Grant No. 2024/20892-8) and CNPq (Grants No. 304891/2022-3 and 40279/2023-8).

\bibliography{bibliography}

\appendix
\makeatletter
\def\@seccntformat#1{%
  \appendixname~\csname the#1\endcsname.\quad
}
\def\@sec@title@format#1#2{#1\quad #2\par}
\makeatother

\clearpage
\widetext

\section{Derivation}\label{sec:app-derivation}

Derivation follows the steps outlined in the Appendix of Ref.~\cite{Rosal2025}.

If we assume that exists a function that evolves 
\begin{equation}
    \vec{y}_{n+1} = \vec{f}_{n+1}(x_{n+1},\vec{y}_n),
\end{equation}
we can write:
\begin{equation}
    \varrho_n(\vec{y}) = \mathrm{E}_{1:n}\left[\rho_n(x_{1:n})\delta_{\vec{y},\vec{y}_n}\right]{,} \quad\quad\quad \rho_{n+1} = \frac{\mathcal{M}_{x_{n+1}}(\vec{y}_n)\rho_n(x_{1:n})}{P(x_{n+1}|x_{1:n})}{.}
\end{equation}

The evolution of the feedback-resolved state is given by
\begin{equation}
\begin{aligned}
    \varrho_{n+1}(\vec{y}) &= \mathrm{E}_{1:n+1}\left[\rho_{n{+1}}(x_{1:n+1})\delta_{\vec{y},\vec{f}_{n+1}(x_{n+1},\vec{y}_n)}\right]{,} \\
    &= \mathrm{E}_{1:n+1}\left[\frac{\mathcal{M}_{x_{n+1}}(\vec{y}_n)\rho_n(x_{1:n})}{P(x_{n+1}|x_{1:n})}\delta_{\vec{y},\vec{f}_{n+1}(x_{n+1},\vec{y}_n)}\right]{.} 
\end{aligned}
\end{equation}

Using the identity $\mathrm{E}_{1:n+1}[\cdot] = \mathrm{E}_{1:n}[\mathrm{E}_{n+1|1:n}[\cdot]]$, we get
\begin{equation}
\begin{aligned}
    \varrho_{n+1}(\vec{y}) &= \mathrm{E}_{1:n}\left[\mathrm{E}_{n+1|1:n}\left[\frac{\mathcal{M}_{x_{n+1}}(\vec{y}_n)\rho_n(x_{1:n})}{P(x_{n+1}|x_{1:n})}\delta_{\vec{y},\vec{f}_{n+1}(x_{n},\vec{y}_n)}\right]\right]{,}  \\
    &= \mathrm{E}_{1:n}\left[\sum_{x'}\mathcal{M}_{x'}(\vec{y}_n)\rho_n(x_{1:n})\delta_{\vec{y},\vec{f}_{n+1}(x',\vec{y}_n)}\right]{.} 
\end{aligned}
\end{equation}

Now we add a summation using a dummy variable $\vec{y}'$:
\begin{equation}
\begin{aligned}
    \varrho_{n+1}(\vec{y}) 
    &= \mathrm{E}_{1:n}\left[\sum_{x'}\sum_{\vec{y}'}\mathcal{M}_{x'}(\vec{y}')\rho_n(x_{1:n})\delta_{\vec{y}',\vec{y}_n}\delta_{\vec{y},\vec{f}_{n+1}(x_{n},\vec{y}')}\right]{,}  \\
    &= \sum_{x',\vec{y}'} \delta_{\vec{y},\vec{f}_{n+1}(x',\vec{y}')} \mathcal{M}_{x'}(\vec{y}') \mathrm{E}_{1:n}\left[\rho_n(x_{1:n})\delta_{\vec{y}',\vec{y}_n} \right]{,}  \\
    &= \sum_{x',\vec{y}'} \delta_{\vec{y},\vec{f}_{n+1}(x',\vec{y}')} \mathcal{M}_{x'}(\vec{y}') \varrho_n(\vec{y}'){.} 
\end{aligned}
\end{equation}

\section{{Normalization and Trace Preservation}}\label{sec:app-trace}
{In this appendix, we show that the feedback-resolved state $\varrho_n(\vec{y})$ remains normalized under the dynamics generated by Eq.~\eqref{eq:deterministic_ME}:
\begin{equation}
\varrho_{n+1}(\vec{y})
=
\sum_{x',\vec{y}'}
\delta_{\vec{y},\,\vec{f}_{n+1}(x',\vec{y}')}
\,\mathcal{M}_{x'}(\vec{y}')
\,\varrho_n(\vec{y}').
\end{equation}
}

{The total state is obtained by summing over all feedback trajectories:
\begin{equation}
\bar{\varrho}_n
=
\sum_{\vec{y}}
\varrho_n(\vec{y})
=
\sum_{\vec{y},x',\vec{y}'}
\delta_{\vec{y},\,\vec{f}_{n+1}(x',\vec{y}')}
\,\mathcal{M}_{x'}(\vec{y}')
\,\varrho_n(\vec{y}').
\end{equation}
}

{Using the identity
\begin{equation}
\sum_{\vec{y}}
\delta_{\vec{y},\,\vec{f}_{n+1}(x',\vec{y}')}
=1,
\end{equation}
we obtain
\begin{equation}
\bar{\varrho}_{n+1}
=
\sum_{x',\vec{y}'}
\mathcal{M}_{x'}(\vec{y}')
\,\varrho_n(\vec{y}').
\end{equation}
}

{Taking the trace,
\begin{equation}
\tr{\,\bar{\varrho}_{n+1}}
=
\sum_{x',\vec{y}'}
\tr{\mathcal{M}_{x'}(\vec{y}')
\,\varrho_n(\vec{y}')}.
\end{equation}
}

{For each feedback configuration, $\vec{y}'$ the measurement map is
trace preserving after summing over outcomes,
\begin{equation}
\sum_{x'}
\tr{\mathcal{M}_{x'}(\vec{y}')\,\rho}
=
\tr{\,\rho},
\end{equation}
it follows that
\begin{equation}
\tr{\bar{\varrho}_{n+1}}
=
\sum_{\vec{y}'}
\tr{\varrho_n(\vec{y}')}
=
\tr{\bar{\varrho}_n}.
\end{equation}
}

{Therefore the total trace is conserved during the evolution,
\begin{equation}
\tr{\bar{\varrho}_n} = 1,
\qquad
\forall n,
\end{equation}
provided the initial state is normalized. Hence the generalized
non-Markovian feedback dynamics defines a trace-preserving evolution.
}

\end{document}